\documentclass[10pt,letterpaper]{article}
\usepackage[top=0.85in,left=2.75in,footskip=0.75in]{geometry}
\usepackage{multirow}

\usepackage{booktabs}
\usepackage{amsmath,amssymb}

\usepackage{changepage}

\usepackage{textcomp,marvosym}

\usepackage{cite}

\usepackage{nameref,hyperref}

\usepackage{microtype}
\DisableLigatures[f]{encoding = *, family = * }

\usepackage[table]{xcolor}

\usepackage{array}

\newcolumntype{+}{!{\vrule width 2pt}}

\newlength\savedwidth

\newcommand\thickhline{\noalign{\global\savedwidth\arrayrulewidth\global\arrayrulewidth 2pt}%
\hline
\noalign{\global\arrayrulewidth\savedwidth}}

\raggedright
\usepackage[aboveskip=1pt,labelfont=bf,labelsep=period,justification=raggedright,singlelinecheck=off]{caption}

\makeatletter
\renewcommand{\@biblabel}[1]{\quad#1.}
\makeatother

\usepackage{lastpage,fancyhdr,graphicx}
\usepackage{epstopdf}
\fancyheadoffset[L]{2.25in}
\fancyfootoffset[L]{2.25in}
\begin{document}
\vspace*{0.2in}

\begin{flushleft}
{\Large
\textbf\newline{Enhancing detection of labor violations in the agricultural sector: A multilevel generalized linear regression model of H-2A violation counts} 
}
\newline
\\

Arezoo Jafari\textsuperscript{1*}, 
Priscila De Azevedo Drummond\textsuperscript{1},
Dominic Nishigaya\textsuperscript{2}, 
Shawn Bhimani\textsuperscript{3}, 
Aidong Adam Ding\textsuperscript{4}, 
Amy Farrell\textsuperscript{2},
Kayse Lee Maass\textsuperscript{1*}
\\
\bigskip
\textbf{1} Department of Mechanical and Industrial Engineering, Northeastern University, Boston, Massachusetts, US
\\
\textbf{2} School of Criminology and Criminal Justice, Northeastern University, Boston, Massachusetts, US
\\
\textbf{3} D’Amore-McKim School of Business, Northeastern University, Boston, Massachusetts, US
\\
\textbf{4} Department of Mathematics, Northeastern University, Boston, Massachusetts, US
\\
\bigskip

* Jafari.a@northeastern.edu, k.maass@northeastern.edu
\end{flushleft}

\section*{Abstract} \label{abstract}
\quad Agricultural workers are essential to the supply chain for our daily food and yet, many face harmful work conditions, including garnished wages, and other labor violations. Workers on H-2A visas are particularly vulnerable due to the precarity of their immigration status being tied to their employer. Although worksite inspections are one mechanism to detect such violations, many labor violations affecting agricultural workers go undetected due to limited inspection resources. In this study, we identify multiple state and industry level factors that correlate with H-2A violations identified by the U.S. Department of Labor Wage and Hour Division using a multilevel zero-inflated negative binomial model. We find that three state-level factors (average farm acreage size, the number of agricultural establishments with less than 20 employees, and higher poverty rates) are correlated with H-2A violations. These findings provide guidance for inspection agencies regarding how to prioritize their limited resources to more effectively inspect agricultural workplaces, thereby improving workplace conditions for H-2A workers. 

\section*{Introduction}
\quad The U.S. agricultural industry is crucial for ensuring food security and contributing to the national economy. This industry comprises various sectors, such as crop production, livestock farming, and food processing. However, it faces numerous challenges such as labor shortages~\cite{martinProposedChangesH2A2022}, climate change~\cite{parajuliEnvironmentalSustainabilityFruit2019}, and a growing demand for sustainable practices~\cite{gollinRoleAgricultureDevelopment2002}. As a result, there is an increasing need for migrant farm workers.

Migrant farm workers are essential for fulfilling the labor demands in the US agricultural sector, particularly in seasonal and labor-intensive areas. International migrants occupy a significant portion of entry-level jobs in agriculture~\cite{martinMigrantWorkersCommercial2016}. Employers often cite the lack of qualified local workers, the superior work ethic of migrants, and, occasionally, the lower labor costs associated with migrant workers as reasons for their preference. This is especially true when migrants are excluded from pension and other benefit programs, which can add 20 to 30 percent to labor costs~\cite{martinMigrantWorkersCommercial2016}. However, migrant workers are susceptible to exploitation, as they face job-related risks, such as exposure to farm chemicals, substandard living conditions, and violations of work hours and wages.

To address labor shortages in the agriculture industry, the U.S. has implemented the H-2 (and, since 1986, the H-2A) visa programs since 1952. These programs enable farmers to fill seasonal farm jobs with guest workers from other countries when they anticipate a shortage of domestic workers, including U.S. citizens, other legally authorized workers, and undocumented workers. Employers seeking to recruit and hire H-2A workers must first obtain certification from the Office of Foreign Labor Certification (OFLC) of the U.S. Department of Labor (DOL)~\cite{martinProposedChangesH2A2022}. It is important to note that the H-2 programs are not new phenomena but rather part of a long history of the United States relying on low-paid workers to support the agriculture industry, beginning with slavery and progressing through sharecropping and the Bracero program, which brought workers from Mexico to fill the labor needs in agriculture.\cite{martinFarmLaborMarket1982, mandeelBraceroProgram194219642014}.

\quad Farm workers on H-2A visas are an essential component of the agricultural industry, as indicated by the rapid increase in H-2A workers employed in the U.S. agricultural industry over the past decade~\cite{castillo2021examining, bruno2020h}. Most agricultural industries have experienced significant growth in H-2A employment, particularly in industries with high labor requirements and seasonal employment~\cite{castillo2021examining}. Migrant Workers with H-2A visas in the United States are limited to working for the employer who applied for their services through the DOL. However, during the COVID-19 pandemic, temporary changes were made to the Department of homeland security requirements for H-2A change of employer requests and H-2A maximum period of stay exceptions~\cite{mizelleTemporaryChangesRequirements2020}. This restriction, coupled with language barriers, lack of knowledge about their rights, and limited access to legal assistance, leaves workers with few options if they encounter abusive work conditions or wage theft. In situations of abusive labor experiences, their only option is to return to their home countries, which they may be reluctant to choose due to losing their job, fear of blacklisting, financial limitations, and other uncertainties. As a result, some workers may feel compelled to continue working under exploitative conditions, leading to workplace abuse~\cite{owens2014understanding}. 

\quad Farm workers, who rely on their employment to support themselves and their families, often face economic vulnerability and hesitate to speak up for themselves, even in situations that may put their lives at risk~\cite{barrickWhenFarmworkersAdvocates2014a}. Additionally, their limited financial resources make it difficult for them to access the legal services necessary to assert their rights fully~\cite{guild2018neighbors}. One of the significant obstacles facing farm workers is their access to health services, as they are exposed to a range of occupational and environmental health risks, which result in high levels of physical injury and illness. The challenges to farm workers' use of health services include inadequate English proficiency, limited education, low income, and a lack of health insurance~\cite{robinson2011wages}.

\quad Several studies discussed strategies to improve and target efforts to protect farm workers~\cite{costa2020federal, Amyk2010,martin2020prosperity}. Farm workers in the United States earn some of the lowest incomes in the labor market and experience a high rate of injuries~\cite{costa2020federal, BLS2020}. The U.S. DOL reveals that the federal and state governments lack the resources for even a one-time inspection of all labor camps in a state~\cite{CDM}. Costa et al. also~\cite{costa2020federal} reported that there is a low probability ($1.1\%$) that each farm employer will be investigated by wage and hour division (WHD) per year. They recommended strategies to target violators, including increasing wage and hour staffing and enforcement funds, targeting the farm labor contractors who are the biggest violators, and ensuring sufficient penalties to stop future violations. Critical to these efforts is understanding factors that increase risk of H-2A violations.

\quad In this paper, we used WHD data that recorded the number of H-2A violations detected by DOL between 2010 and 2020. To examine the factors influencing the number of H-2A violations, we assembled a dataset containing various factors from multiple sources. This allowed us to pinpoint the factors significantly affecting H-2A violations, thereby offering insights for optimizing resource allocation, such as budget, time, and personnel. Furthermore, the study aimed to prioritize worksites based on the likelihood of labor violations, thereby improving the effectiveness of strategies to address such violations. We employed a multilevel zero-inflated negative binomial model to identify the correlation between factors and the count outcome. Our findings provide valuable insights that can be used by inspection agencies to optimize their investigative strategies, resulting in more efficient and effective enforcement of labor regulations.

\quad  Our study makes several contributions to the literature on H-2A violations in the agricultural industry.  Firstly, we conducted a literature review to identify the key factors that may influence the detection of H-2A violations. Secondly, while previous studies have used quantitative analyses on certain factors, our study is unique in its application of a multilevel zero-inflated negative binomial model to the best of our knowledge. This approach allows for exploring correlations between significant factors and H-2A violations at both state and industry levels, providing a better understanding of the issue. Thirdly, based on the suggested model, we found that the average size of farms and number of small/medium establishments are correlated with the mean of H-2A violations. Additionally, labor intensity, number of task forces, duration of alleged violations, fatal injury rate, and minimum standard wage are associated with the likelihood of zero H-2A violation counts at both state and industry levels. The poverty rate is associated with both mean of H-2A violations and the likelihood of zero H-2A violation counts. By pinpointing the significant factors associated with H-2A violations, our study offers valuable insights that can guide policy decisions in the realm of labor regulation and enforcement.

\quad The rest of this paper is structured as follows: The next Section provides a literature review on factors that may influence H-2A violations and discusses relevant multilevel and zero-inflated models employed in previous studies. The Data Section details the factors considered for our regression analysis and outlines the data sources used to obtain these variables. In the Method Section, we describe the regression models and multilevel generalized linear models applied to identify correlations between factors and H-2A violation counts. The Result Section presents and interprets our analysis results, emphasizing key factors that significantly correlate with H-2A violations in the agricultural industry. Lastly, in the Discussion and Conclusion Sections, we summarize our findings, discuss the practical implications of our study, and suggest strategies to improve the effectiveness of investigations. Overall, our research provides valuable insights that can inform policy decisions and support the development of more effective approaches for detecting labor violations in this industry.

\section*{Literature Review} \label{LR}
\quad This section aims to briefly review  relevant studies focused on factors influencing the risk of labor violations, as well as regression models employed in the analysis of similar data structures.
\subsection*{Factors Affecting Labor Violations Among Agricultural Workers} \label{LR on factors}

\quad Many scholars have studied the factors that increase the vulnerability of farm workers to exploitation. Agricultural work environments are notorious for increased risk of occupational injuries due to environmental heat stress, pesticide exposure, heavy workloads, and other workplace safety concerns~\cite{Zhang324, mac2017farmworker, vega2019impacts, STOKLOSA2020215, robinson2011wages, arcury2015job, ronda-perezLabourTraffickingChallenges2017}, including fatal injuries ~\cite{pratt1998injury,pickett1999fatal, Horsburgh489, rivara1985fatal}. Social distancing was particularly challenging in agricultural workplaces during the COVID-19 pandemic~\cite{luckstead2021labor}. Prior studies also document the impacts such work environments and stressors have on farm workers' mental health~\cite{hiott2008migrant,safety2040023, earle2003occupational, lindert2009depression}. Additionally, wage violations~\cite{marinescu2021wage, nagurney2022attracting, Schwarz2019} and violence~\cite{norwood2020labor} are widely documented in the agricultural industry. Prior research has explored how these issues are impacted by the contract type and recruitment strategy~\cite{silverman2016troubling, costa2020federal, castillo2021examining, siqueira2014effects}, farm size~\cite{Bhorat2012, pierre2013firms, harrisonFarmSizeJob2015}, demographic characteristics of labors~\cite{barrick2014labor, zhang2014estimating, richards2004trafficking}, migrant workers~\cite{CDM, castillo2021examining} and labor inspections~\cite{barrick2014farmworkers, farrell2020policing}. We sought to explore the relationship of these factors to the number of H-2A violations detected by the DOL. However, due to the lack of available data on each of these factors during a common time period, our analysis focuses on the subset of factors for which data is available (as described further in the Data Section). Below we summarize the literature on the aforementioned factors for which data is available and are included in our study. Specifically, we focus on farm size (according to multiple metrics: acreage, number of employees), labor intensity, average net income per farm, poverty rate, and fatal injury rate. We also explore the relationship between the presence of human trafficking taskforces and the state minimum standard wage on detected H-2A violations.  

\quad \textit{Farm size: }Several studies have focused on farm size in relation to labor exploitation. Some have found that labor regulations are less strict for smaller firms (number of employees)~\cite{Bhorat2012, pierre2013firms}. An article examining the impacts of firms' size on the risk of labor violations found a high probability of labor violation among employees in small (less than ten employees) and medium-sized firms (10 to 19 employees) rather than larger enterprises~\cite{Bhorat2012}.

\quad The literature highlights the importance of considering the correlation between the land size of farms (measured in acres) and labor violations, in addition to the impacts of labor laws on farms with fewer employees. Harrison and Getz~\cite{harrisonFarmSizeJob2015} found that although larger acreage farms generally offer better job quality than smaller ones, these advantages are disproportionately available to white, U.S.-born workers. Migrant workers face challenges such as job insecurity, limited professional growth, and fear of losing their jobs due to their legal status. The study highlights the need to examine the relationship between farm size and labor violations, considering the varying job quality and opportunities for different worker groups. Larger acreage farms may not always be more prone to labor violations, but specific subgroups like migrant workers could be more vulnerable due to the mentioned factors.


\quad \textit{Labor intensity: }The main characteristic of high-intensive jobs is the low use of machinery and intensive use of manual labor \cite{Zahniser2018}. These increase workers' exposure to nature and its adversities. Articles have also documented that high-intensive work is highly dependent on temporary migrant workers~\cite{castillo2021examining, moseley2006working}. One such group of migrant workers that is particularly relevant to this research is workers on H-2A (agricultural) visas. Castillo et al.~\cite{castillo2021examining} described the trends in the H-2A program in the agriculture sector, finding  H-2A employment growth in vegetables and melons, fruits, and tree nuts that reflect sectors with higher labor intensity. Since migrant workers are susceptible to labor exploitation because of their apprehension about job loss, heightened exposure to hazardous work environments, and extended working hours~\cite{preibischDoesCitizenshipStatus2014}, it is crucial to investigate H-2A violations in labor-intensive workplaces. 

\quad \textit{Income: }The literature highlights the need to examine the correlation between farm income and labor exploitation in the agricultural sector. We could not find articles that specifically identify a relationship between farm income and labor violations in the agricultural industry. However, the literature discussed that businesses maximize labor efficiency to earn profits. This pressure to extract maximum profit from the work process often results in various labor standard issues, such as excessive working hours, low wages, and an over-reliance on contingent migrant workers\cite{lockeWorkplaceUpstreamBusiness2018}.

\quad \textit{Task forces: }Barrick et al.~\cite{barrick2014farmworkers} studied law enforcement's ability to identify, investigate and prosecute labor trafficking among farm workers. The study revealed a discrepancy between the perceptions of law enforcement, including task forces, regarding labor trafficking operations and the actual reports of labor trafficking by agricultural workers in parts of North Carolina. They recommended that law enforcement agencies in agricultural areas extend their missions to include the protection of laborers, farms, and farm camps in their routine investigations. Farrell et al.~\cite{farrell2020policing} explored strategies law enforcement, including multi-agency task forces, can use to improve detection, developing non-traditional partnerships with labor inspections and local regulatory agencies.

\quad \textit{Poverty rate: }Schwarz et al.~\cite{Schwarz2019} presented major clusters of trafficking risk factors, including economic insecurity, house insecurity, education gaps, and migration. They expressed that the relationship between poverty and labor trafficking is strong since poverty restricts people's options and makes them vulnerable to labor exploitation. Also, several studies discussed that the poverty rate influences the chance of children becoming child laborers~\cite {refId0}. Marinescu et al.~\cite{marinescu2021wage} identified the negative correlation between wages and violations enforced by OSHA and WHD. 
Nagurney~\cite{nagurney2022attracting} employed a network equilibrium model to demonstrate that wages play a crucial role in decisions to hire migrant workers. The study also revealed that engaging in illicit practices, such as misleading potential workers about their compensation, can financially benefit farmers, thereby emphasizing the need for stricter oversight and control. 

\textit{Fatal injury rate: }Some articles suggest that agriculture is one of the most dangerous occupational sectors regarding fatal injuries~\cite{pratt1998injury,pickett1999fatal, Horsburgh489}. Rivara~\cite{rivara1985fatal} studied fatal and nonfatal injuries among agricultural workers in the US and found that farm machinery, such as tractors, is the most common reason for fatal and nonfatal injuries. 

\quad \textit{State minimum standard wages: }The Fair Labor Standards Act (FLSA) establishes the federal minimum wage and directly elicits the minimum wage provisions for employers to compensate all employees legally. The literature suggests that when the minimum wage increases beyond what a firm can or is willing to pay, maintaining a job match may necessitate paying subminimum wages to some workers~\cite{clemensUnderstandingWageTheft2022}. As a result, workers might stop seeking enforcement of the minimum wage, for instance, by choosing not to report violations of minimum wage regulations when they feared their job is in danger~\cite{yanivMinimumWageNoncompliance2001, clemensUnderstandingWageTheft2022}. It is important to consider the role of minimum wage standards in labor violations, particularly within the agricultural sector. 

\subsection*{Multilevel and Zero-inflated Models} \label{LR on modeling}
Multilevel modeling is a robust statistical technique for analyzing data with group structures. This method has been employed across various disciplines to tackle diverse research questions, effectively demonstrating its worth in discerning both within-group and between-group effects and interactions~\cite{garson2019multilevel}.
Multilevel modeling enables researchers to accommodate intricate data structures and has found applications in various social science domains, such as education~\cite{khine2022methodology, shirilla2022benefits}, health~\cite{leyland2020multilevel, seidu2020barriers}, environmental studies~\cite{paul2017multilevel, ECHAVARREN2019813}, and violence-related research~\cite{UTHMAN20091801, KISS20121172}.

\quad  Some articles have depicted the linear association between predictors and continuous outcomes by constructing multilevel linear models for detecting significant factors~\cite{Leonard2016, ctx16693458590001401}. On the other hand, the generalized linear multilevel models typically utilized for analyzing count, binary or categorical data differ from the former. Leclerc et al.~\cite{Leclerc} employed a mixed-effect logistic regression analysis to study the effects of potential guardianship on the severity of child sexual abuse for nested data. Martinez-Schuldt et al.~\cite{Ricardo2021} studied the willingness of immigrant community members to notify law enforcement after being victimized. They used multilevel logistic regression analysis for a binary dependent variable. 

\quad Several studies particularly focused on count data when the data are clustered or grouped~\cite{macdonald2010count, HUEBNER2003107, Marchment2021,shaaban2021statistical, almasi2016multilevel, HIDANO201965, seabright2023repercussions}, using Poisson distribution~\cite{HUEBNER2003107, Marchment2021} or negative binomial distribution~\cite{shaaban2021statistical}. Count data often exhibit excess zeros, meaning there are more zeros in the data than expected from a Poisson or negative binomial distribution. In such cases, two commonly used models are the zero-inflated Poisson (ZIP) regression model and the zero-inflated negative binomial (ZINB) regression model~\cite{almasi2016multilevel, HIDANO201965, seabright2023repercussions, rorie2022structural, Niedhammer2012, forst2015spatial}.
Seabright et al.~\cite{ seabright2023repercussions} fitted zero-inflated generalized linear multilevel models to investigate the potential impacts of  post-marital residence patterns on the size of women’s social groups, as well as their access to alloparental childcare among Tsimane fora-ger–farmers in lowland Bolivia. An article presented a ZINB mixed effect model to identify predictors' effects on adolescent victimization~\cite{rorie2022structural}. Niedhammer et al.~\cite{Niedhammer2012} examined the association between psychosocial work factors and sickness absence in 31 Europe countries. They used the multilevel negative binomial hurdle model to study the count of sickness absence with three hierarchical levels and excessive zeros. Using the zero-inflated negative binomial model with random effects, Forst et al.~\cite{forst2015spatial} modeled the association between factors, including the occupational category and demographic characteristics, and work-related injury counts for repeated measures within zip codes. This study provided evidence to support the potential benefits of community-based approaches for reducing the burden of workplace injuries and promoting occupational health and safety.

\quad Our study employs multilevel and zero-inflated models as valuable tools for understanding the factors associated with H-2A labor violations. Leveraging a unique dataset, our analysis contributes to the existing literature on the factors influencing H-2A labor violations. In addition, multilevel modeling is particularly advantageous for our research, as it accounts for varying predictor effects across categories, such as state or industry. This observed variation in our data allows for a more accurate assessment of the relationship between these factors and H-2A violation counts compared to previous methods.
Furthermore, we incorporate a zero-inflated model in our study to address the presence of structural zeros in the data. This is essential for our analysis, as there are likely two distinct mechanisms generating detected H-2A violations, one of which does not produce detected violations greater than zero count. The zero-inflated model accurately accounts for these structural zeros, offering a more comprehensive understanding of the factors influencing H-2A labor violations.

\section*{Data}\label{data}
\quad The data used in this analysis comes from multiple sources, which are described in detail in this section. We also discuss the data cleaning process and feature selection to identify the most important factors used in the regression analysis. 

\subsection*{Data Collection} \label{data collection}
\quad This research utilizes data from multiple sources to examine the correlation between various factors and H-2A visa program violations, which could help inform policymakers and inspection agencies in detecting violations in the agricultural industry. To analyze labor violations in the agricultural sector, we used data from the Wage and Hour Division (WHD) of the U.S. Department of Labor (DOL) covering the period from 2010 to 2020~\cite{WHD2023}. Specifically, cases with a ``findings start date'' and ``findings end date'' between 2010 and 2020 were included. The dataset includes information on multiple types of violations, such as H-2A, H-1A, H-1B, Occupational Safety and Health Administration, and Fair Labor Standards Act violations, per case, investigated by DOL. It includes 175,126 cases investigated by DOL in a wide range of industries. Given our research's focus on labor violations in the agricultural sector, we relied on a specific subset of data related to the count of H-2A violations per case. Specifically, the data was filtered to include only those cases with North American Industry Classification System (NAICS) codes that began with 11, corresponding to agriculture, forestry, fishing, and hunting industries. This filtering ensured that the analysis focused specifically on labor violations in the agricultural sector where H-2A violations may occur and resulted in 12,041 remaining cases. We next excluded H-2A violation counts for Puerto Rico and the District of Columbia from our analysis as our study focused on the 50 U.S. states. Additionally, finding data on relevant factors for these regions was challenging. This further reduced the number of cases from 12,041 to 11,976. Of the 11,976 cases within the 50 U.S. states with NAICS code starting with ``11", 2,523 were found to have greater than zero H-2A violations reported. However, it is important to note that for the cases with zero H-2A violations reported, it is unclear whether the DOL investigated for H-2A violations and found no violation occurred or whether DOL focused on investigating other types of labor violations and therefore wasn't looking for H-2A violations. This presented a challenge for the analyses, and we explained in detail how we addressed it in the methodology section. Additionally, some cases were found to have multiple H-2A violations. Thus, the 2,523 cases with non-zero H-2A violations had a total of 73,781 H-2A violations. 

\begin{figure}[!h]
\begin{center}
        \includegraphics[width=1\textwidth]{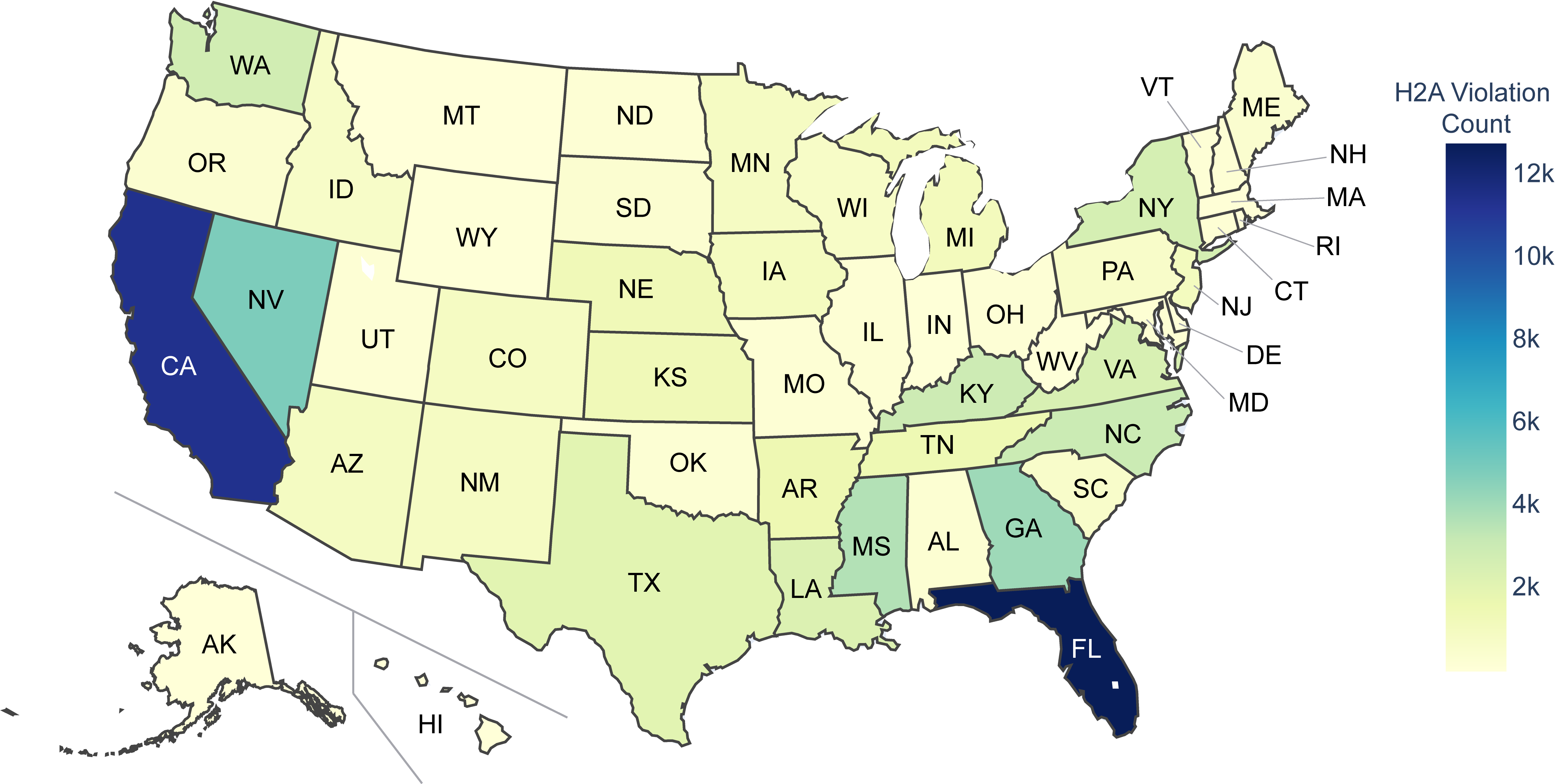}
\caption{{\bf Geographical Distribution of H-2A Violations.} The heatmap illustrates the total number of identified H-2A violations per state from 2010 - 2020.}
\label{fig1}
\end{center}
\end{figure}
\begin{figure}[!h]
\begin{center}
        \includegraphics[width=1\textwidth]{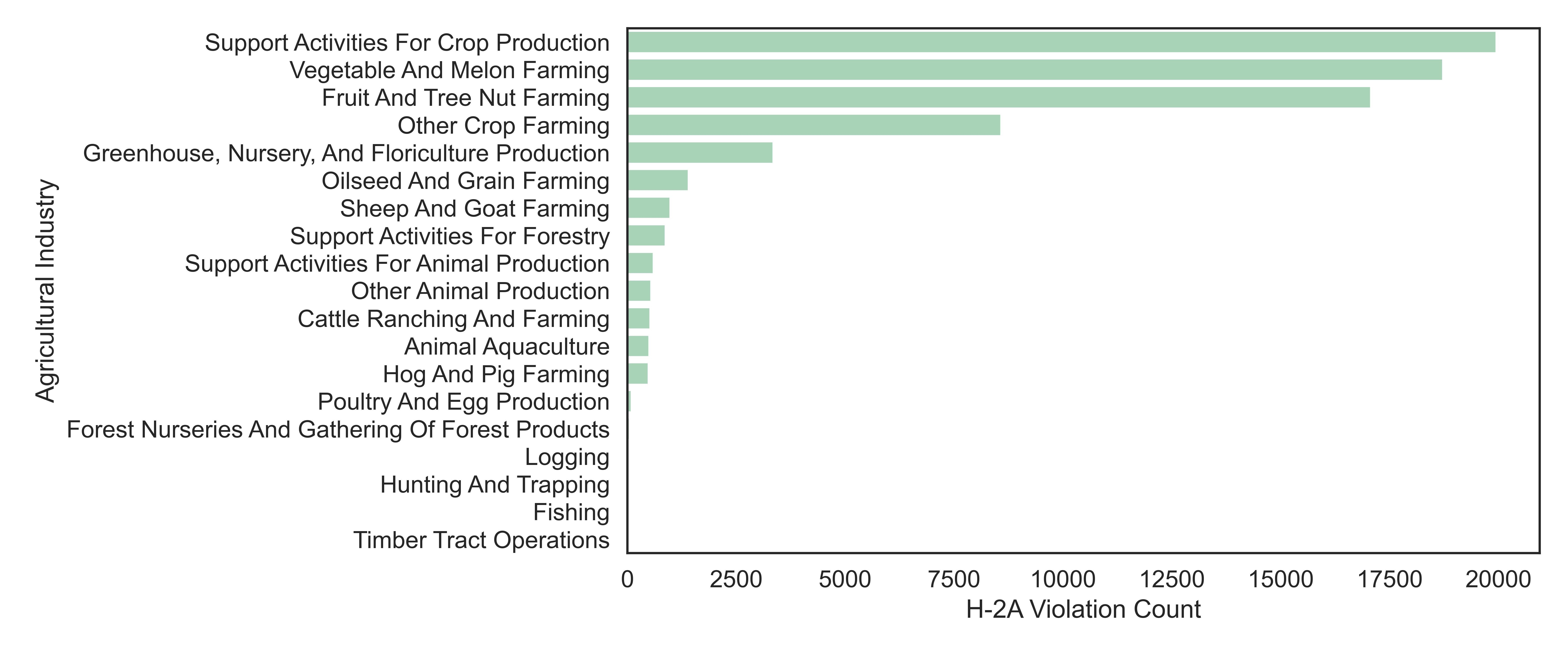}
\caption{{\bf H-2A Violations by Industry.} The bar plot demonstrates the distribution of H-2A violation counts across agricultural industries, highlighting the industries with the most violations from 2010 - 2020.}
\label{fig2}
\end{center}
\end{figure}

\quad H-2A violations were found in all 50 states and 17 of the 19 agricultural industries over the 10 year period analyzed (see Figures \ref{fig1} and \ref{fig2}, respectively). Florida, California, and Nevada clearly show more H-2A violations than other states. However, it is worth noting that factors such as the number of farms, climate, and  population of these states may contribute to the higher number of violations and that the heatmap provides the count of H-2A violations identified without normalizing for these factors. The bar chart reveals that substantially more H-2A violations have been identified in the Support Activities for Crop Production (19,950), Vegetable and Melon Farming (18,727), and Fruit and Tree Nut Farming (17,073) industries than in other agricultural industries; some industries have had very few H-2A violations identified, such as Poultry and Egg Production (96), Forest Nurseries and Gathering of Forest Products (52), Logging (14), and Hunting and Trapping (2). No H-2A Violations were identified in the Fishing and Timer Tract Operations industries during the 10 year period. The combined analysis highlights geographical and industry-specific H-2A violation trends.



\quad We combined WHD data with external data from multiple sources to form our consolidated dataset. The external data consists of various factors that we categorized into three groups: agriculture land, inspection, and population, see Table~\ref{table1}. Several studies in the Literature Review Section which discussed the potential factors affecting the vulnerability of farm workers motivated us to study their impacts on the number of H-2A violations. In the following paragraphs, we describe the factors considered in the analysis.

\begin{table}[!ht]
\caption{{\bf Data Frame}}
\begin{tabular}{l|l}
\hline
\bf Category &\bf Factor  \\ 
\thickhline
\multirow{6}{*}{Agriculture Land}
& Average Size of Farms (acres)\\ & Number of Farms\\ & Number of Small/Medium Establishments\\ & Labor Intensity\\ & Average Net Income per Farm\\
\specialrule{0.001pt}{2pt}{2pt}
\multirow{2}{*}{Inspection} & Number of Task Forces \\ & Length of Alleged Violations\\ 
\specialrule{0.001pt}{2pt}{2pt}
\multirow{3}{*}{Population} & Poverty Rate \\ 
& Fatal Injury Rate \\ & Minimum Standard Wages\\
\hline
\end{tabular}
\label{table1}
\end{table}

\quad  \textit{Average size of farms (acres)} --
 Each state's average acreage size of farms was collected from the United States Department of Agriculture (USDA) database from the National Agricultural Statistics Service (NASS)~\cite{FarmLandUSDA} using the \texttt{Average farm size (acres)} field. The data represent the size of
any place from which a thousand dollars or more of agricultural products were produced, sold, or normally would have been sold during the fiscal year. We collected data from 2010 to 2020. This dataset contained information for the number of farms, land in farms (acres), and average farm size per state annually. The average farm size ranged from 55 (Rhode Island) to 2745 acres (Wyoming). 

\quad  \textit{Number of farms} -- 
The number of farms in each state from 2010 to 2020 was obtained from the NASS USDA database field \texttt{Number of farms} field~\cite{FarmLandUSDA}. The number of farms per state ranged from 680 farms (Alaska, 2010-2012) to 248,500 farms (Texas, 2013 \& 2017), with an average number of 41,653 farms per state.


\quad  \textit{Number of small/medium establishments (by number of employees)} --
The number of small/medium
establishments was collected from the Quarterly Census of Employment and Wages
(QCEW) database from the U.S. Bureau of Labor Statistics ~\cite{QCEWdataset2022}. The data represents
the employment data from companies in the U.S. by type of industry (NAICS code) and
state. It represents the number of establishments recorded quarterly in each year. The data available is only for the number of farms counted in the first quarter of each year. It might exclude H2A workers depending on the state labor regulation, which is a limitation. Although this data is expected to exclude around 20 \% of the agricultural workers, it is considered a representative dataset for operation in the
agricultural sector from 2010 to 2020 used in USDA reports~\cite{castillo2021examining}. We define small/medium farms as those with less than 20 employees. The justification for this threshold is based on regulations on labor rights. For example, under California law, organizations with less than 26 employers abide by less strict labor regulations, being allowed to pay lower minimum wage~\cite{MinimumWageFrequently}. The U.S. Federal legislation exempts the minimum wage for certain organizations. Specifically, organizations with an annual gross volume of sales or business done of at least \$500,000 and specific organizations with fewer than nine employees are exempt from this requirement~\cite{MinWageFED_law}. Additionally, employees who work for companies with fewer than ten employees do not have the right to receive unemployment insurance~\cite {DOL_UI}. The dataset includes different categories for firm sizes, such as less than 5 employees, 5 to 9 employees, 10 to 19 employees, 20 to 49 employees and other categories. The dataset comprises various categories for firm sizes, including less than 5 employees, 5 to 9 employees, 10 to 19 employees, 20 to 49 employees, and other categories. As a result, we combined the groups with fewer than 20 employees to represent the small/medium establishment variable, as small/medium-sized firms with less than 20 employees have been defined in a previous study~\cite{Bhorat2012}.

\quad \textit{Labor intensity} --
To study the effects of labor intensity on labor violation counts, we classified each type of crop in the NAICS code as either high-intensity (1) or low-intensity manual labor (0). A labor-intensive crop is a relative measure of the number of hours of human labor required to produce the same yield in dollars in comparison to other crops~\cite{castillo2021examining}. It was difficult to establish a definitive threshold to determine which crops require high-intensity labor due to factors such as technological advancements and changes in farming practices over time~\cite{huffman2005trends}. Therefore, we used the same classification as Castillo et al.~\cite{castillo2021examining} used in a 2021 USDA report. According to their findings, high-intensive crops usually employ more H-2A workers than non-high-intensive ones~\cite{castillo2021examining}. To help improve the classification, we also used other papers to corroborate the classification of which NAICS code is associated with high-intensity labor~\cite{costa2020federal, huffman2005trends}.

\quad \textit{Average net income} --
Net farm income was extracted from the U.S. Department of Agriculture (USDA), Economic Research Service, Farm Income and Wealth Statistics~\cite{FarmSectorFinancial}. This dataset includes the \texttt{net farm income} per state by year for the range of 2010 - 2020. To find the average net income per farm, the net farm income is divided by the number of farms per state by the year. The number of farms is gathered by USDA's reports of farms and land in farms~\cite{FarmLandUSDA}. The average yearly net income varied between -\$21,369.12 to \$263,372.17 across states.

\quad \textit{Number of task forces} --
Due to the variety of human trafficking task forces throughout the US and a lack of data related to how many task forces exist in each state, we focused on the number of enhanced collaborative model task forces funded through the Office of Justice Programs Office for Victims of Crime (OVC) that aim to combat human trafficking~\cite{OJP}. These task forces include victims, law enforcement, social service providers, and other governmental and non-governmental partners to help trafficking victims and provide them with appropriate services.  ECM task forces have specific provisions focused on labor trafficking which may provide the best indicator of law enforcement awareness and readiness to identify labor trafficking and associated exploitation. This data consists of information about the funded task forces, award amount, award status (e.g., open, closed, etc.), etc. To transform the data for regression analysis, we counted the unique number of active OVC task forces per by year. The number of OVC funded task forces ranged from 0 to 15 (California, 2019), with an average of 1.002 task forces in a state each year. 

\quad \textit{Length of alleged violations} --
We calculated the length of the alleged violations data per case investigated in the WHD dataset by taking the difference of the \texttt{findings\_end\_date}  and the \texttt{findings\_start\_date} of the alleged violations~\cite{WHD2023}. This time period indicates the estimated duration of violations that occurred per case. Only cases related to agricultural businesses (filtered by NAICS codes that begin with 11) during 2010 - 2020 were included in our analysis. To determine the state associated with the business in each case, we used the \texttt{st\_cd} field in the WHD dataset.

\quad \textit{Poverty rate} -- We obtained poverty statistics by state from The U.S. Census Bureau historical poverty tables~\cite{bureauHistoricalPovertyTables}. Specifically, we used data from the \texttt{Percent in Poverty} field in \textit{Table 19. The number of Poor and Poverty Rate by State: 1980 to 2021} for the years 2010 to 2020. Although we study labor violations in agriculture,  due to the lack of available data related to poverty rates specifically for people employed in the agricultural sector, the poverty rate data we used includes households working in agricultural and non-agricultural jobs. The data ranges from a minimum poverty rate of 0.9\% (New Hampshire, 2009-2010) to 25.8\% (North Dakota, 2012).

\quad \textit{Fatal injury rate} -- As the lack of tracking systems makes detecting and understanding trafficked workers' health and risk patterns difficult, we extract data from the U.S. Bureau of Labor Statistics, injuries, illness, and fatalities (BLSIIF)~\cite{BLS2020}. According to the data's resources, it can be used to compare risk among worker groups with varying employment levels. Due to incomplete data on nonfatal injuries during our specified timeframe, we relied on the fatal injury rate for our analysis. The data consists of fatal rates across states from 2007 to 2021. However, we used data ranging from 2010 to 2020, which is within the scope of our analysis. The data ranges from a minimum fatal injury rate of 0.9\% (New Hampshire, 2009-2010) to 17.7\% (North Dakota, 2012). 

\quad  \textit{Minimum standard wages} --
We used the DOL wage data for changes in basic minimum wages under state law from 2010 to 2020 obtained by the U.S. DOL Division of Fair Labor Standards Act and Child Labor Wage and Hour Division~\cite{MinWageDOL}. DOL wage data has some exceptions related to minimum wage values for specific states. To maintain consistency and uniformity in data analysis, we addressed issues by taking the following steps; First, data includes minimum wage values for some states that are lower than the federal minimum wage due to some exceptions, such as a low number of employees. We decided to use the federal minimum wage for states rather than the lowest wage a state can legally pay if all exceptions are met. Second, five states do not have a state minimum wage, which leaves them subject to federal law. We used the federal minimum wage of \$7.25 for these states for our analysis~\cite{NCSL}. Third, some states have a range of numbers for the minimum wage, indicating that the pay could be anywhere between two numbers. In these cases, we decided to conservatively use the higher value since, in some states, this may indicate the state’s initiative to increase wages within their legal framework.

\begin{table}[!ht]
\caption{{\bf Description of factors.}}
\begin{tabular}{llll}
\hline
\bf Factor &\bf Source & \bf Type & \bf State/Industry \\ 
\hline
\text{Average Size of  Farms} & \href{https://usda.library.cornell.edu/concern/publications/5712m6524}{USDA}  & integer & state\\ 
\text{Number of Farms} & \href{https://usda.library.cornell.edu/concern/publications/5712m6524}{USDA}  & integer & state\\ 
\text{Number of Small/Med Establishments} & \href{https://www.bls.gov/cew/downloadable-data-files.htm}{QCEW} & integer & state\\ 
\text{Labor Intensity} & USDA,~\citenum{castillo2021examining}  & binary & industry\\ 
\text{Average Net Income per Farm} & \href{https://data.ers.usda.gov/reports.aspx?ID=17839#Pc18b4a9a90a54f44becbae9df249f5c8_6_185iT0R0x16}{USDA} & continuous & state\\ 
\text{Number of Task Forces} & \href{https://data.ojp.usdoj.gov/Funding-API/Funding/a7v8-ei2f}{OJP} & integer & state\\ 
\text{Length of Alleged Violations} & \href{https://enforcedata.dol.gov/views/data_catalogs.php}{WHD} & integer & state and industry\\ 
\text{Poverty Rate} & \href{https://www.census.gov/data/tables/time-series/demo/income-poverty/historical-poverty-people.html}{USCB} & continuous & state \\ 
\text{Fatal Injury Rate} & \href{https://www.bls.gov/iif/state-data/fatal-injury-rates-by-state-2007-2021.htm}{BLSIIF} & continuous & state \\ 
\text{Minimum Standard Wages} & \href{https://www.dol.gov/agencies/whd/state/minimum-wage/history}{DOL} & continuous & state \\ 
\hline
\end{tabular}
\label{table2}
\end{table}

\subsection*{Data Pre-processing and Feature Selection}\label{data preprocessing}

\quad In this section, we first applied feature scaling to standardize all regressors, as they were initially on different scales. This process ensured that no particular predictor would dominate the analysis due to its scale. To explore the dataset further and identify correlations between regressors or factors, we utilized a correlation matrix, which provided valuable insights into the relationships among the variables.

To check for multicollinearity, we employed the variance inflation factor (VIF) to assess the degree of independence among the factors. We discovered that some factors, such as the number of farms, number of small/medium establishments, and average net income per farm, exhibited high VIF values. To select the most relevant features, we employed Pearson correlation analysis and found that the number of small/medium establishments had the strongest correlation with the target variable, outperforming the other two factors. Additionally, we leveraged Feature Importance from the random forest regression model and used recursive feature elimination with k-fold cross-validation to determine the most influential factors for our model. These methods, including the correlation matrix and random forest feature importance, indicated that the number of small/medium establishments should be considered as a factor in our model.

Lastly, we had to select between the number of farms and the average net income per farm based on the VIF results. Since the average net income per farm was derived from the number of farms, we decided to retain the average net income variable, as it implicitly represents the number of farms. This choice allowed us to avoid potential multicollinearity issues while still incorporating valuable information in our analysis. All data cleaning and pre-processing tasks were performed using Python 3.7, ensuring a consistent and robust approach to preparing the data for subsequent analysis. This streamlined process allowed us to efficiently identify and incorporate the most relevant factors into our model, ultimately yielding more accurate and meaningful results.
\section*{Method} \label{method}
\quad In this section, we will explain the modeling approach used to establish a relationship between the predictors and the H-2A violation count, which is the target variable. We begin by exploring various regression models suitable for count data, where many observations have zero counts. Subsequently, we will implement a multilevel model for the grouped data with two levels, state and industry. 

\subsection*{Regression Models For Count Data}
\quad The dependent variable is the number of H-2A violations detected by DOL per case. It can only take non-negative integer values and experiences excess zeros. These types of count data are modeled with General linear models (GLMs) and General linear mixed models (GLMMs) using either negative binomial (NB) or Poisson distributions~\cite{mwalili2008zero,perumean2013zero}. 

\quad We first investigated whether the Poisson distribution would be an appropriate fit for our data by conducting a dispersion test. Upon examining the results, we found that the p-value was statistically significant, indicating the presence of overdispersion in the count data. This overdispersion suggests that the variance is greater than the mean of the outcome variable, which is not ideal for a Poisson distribution. We decided to consider the NB regression model since this model is more adept at handling overdispersed count data, making it a more suitable choice compared to the Poisson regression model.
It is common to use a zero-inflated model when there are more zeros in count data than a simple model predicts. To check whether such zero-inflated models are needed, we conduct a model comparison between the simple models and zero-inflated models. Zero-inflated general linear models are a mixture of distributions, one that has degenerate zero counts (logit)\textemdash a zero-inflated model\textemdash and another that has degenerate integer counts following Poisson and Negative binomial (NB) distributions\textemdash a conditional model~\cite{Lambert1992}.
We compared Zero-inflated Negative Binomial (ZINB) and NB models fitted to the dataset using the Akaike information criterion (AIC), Bayesian information criterion (BIC), and likelihood ratio test. AIC and BIC serve as measures for assessing the goodness of fit of statistical models while considering their complexity. Both criteria aim to find the optimal balance between fitting the data well and avoiding overfitting due to unnecessary complexity. Lower values of AIC and BIC signify a better fit of the model to the data, as they indicate a more favorable balance between data fit and model complexity. The formula used to compute AIC is $AIC = -2log(L) + k$, where $L$ is the maximum value of the likelihood function for the model and $k$ is the number of estimated parameters in the model~\cite{Cavanaugh2019}. The BIC penalizes the model complexity more strongly than the AIC, and both are useful for model selection, depending on the trade-off between goodness of fit and model complexity. As shown in Table~\ref{table3}, the ZINB model has a lower AIC and BIC value compared to the NB model, indicating the zero-inflated component of the model effectively captures the remaining overdispersion that the NB model does not entirely explain. In addition to AIC and BIC, we used the likelihood ratio test to compare the two models. A likelihood ratio test is a validation approach that compares a simple model, such as the NB model, with a special case of that model that includes the zero-inflated part, such as the ZINB model. According to the test, additional parameters in the ZINB model improve the model's fit significantly compared to the NB model, confirming the conclusion of the AIC and BIC comparison. Therefore, we used ZINB regression for the number of H-2A violations that exhibit overdispersion and excess zeros. 
\begin{table}[!ht]
\caption{{\bf Model selection}}
\begin{tabular}{l|ll}
\hline
\bf Model &\bf AIC & \bf BIC  \\ 
\thickhline
$NB$ $Model$ & 30184.2 & 30265.5\\ 
$ZINB$ $Model$ & 28509.4 & 28664.6\\ \hline
\end{tabular}
\label{table3}
\end{table}

\quad The fitted ZINB model reflects two possible outcomes for H-2A violations reported in this dataset: zero H-2A violations or a positive number of H-2A violations. Some cases reporting zero H-2A violations reflect structural zeros, while others reflect non-structural (random) zeros. Structural zeros encompass cases in which a zero is reported in the dataset because either (a) DOL investigated other (non-H-2A) types of violations; since they were not looking for H-2A violations, no violations were found or reported, or (b) the organization's structure is such that H-2A violations are not possible, such as the organization does not hire H-2A workers. In investigation cases where DOL is inspecting for H-2A violations and the organization's structure is such that a H-2A violation could potentially occur, the reported number of H-2A violations can be zero or greater than zero and are generated using the NB distribution portion of the model. Zero counts in these cases are non-structural (random) zeros and refer to situations in which the organization could potentially have had violations, but no violation was found because even though DOL investigated for H-2A violations, (a) the organization did not have a violation during the period in which they were investigated or (i.e., true negative) (b) DOL could not detect the H-2A violations that were occurring during the inspection (low detectability) (i.e., false negative).

We assumed that structural zeros occur with probability $\Phi$. Therefore, the probability mass function (pmf) for a ZINB  can be formulated as follows~\cite{Young2022}:

\begin{eqnarray}
\label{eq:schemeP}
	\mathrm{Pr(Y = y)} =\begin{cases}
      \Phi + (1 - \Phi) g(Y=0), & \text{if}\ y=0 \\
      (1 - \Phi) g(Y=y), & \text{if}\ y = \mathbb{N}^+, 
    \end{cases}
\end{eqnarray}
 \quad where the random variable  $Y\in \mathbb{N}$ is the count data, $y$ is the realization of $Y$, and $g(Y)$ is the pmf of the negative binomial distribution. The distribution is formulated in terms of the mean ($\mu$) and dispersion parameter ($\alpha = \frac{1}{r}$), where $r$ is the predefined number of successes that occur according to the definition of the NB distribution. 
\begin{eqnarray} 
\label{eq:schemeP2}
	{g(Y = y) = Pr(y|\mu,\alpha)} =\frac{\Gamma(\alpha^{-1}+y)}{\Gamma(y + 1) \Gamma(\alpha^{-1})} \left( \frac{1}{1+\alpha \mu}\right)^{\alpha^{-1}} \left( \frac{\alpha \mu}{1+\alpha \mu}\right)^{y}. 
\end{eqnarray}
\quad The negative binomial components are used to estimate the intercept and coefficients of the regression model fitted to the dataset, thereby allowing us to accurately capture the relationships between the variables and make meaningful inferences based on the fitted model. 

\subsection*{ Multilevel Generalized Linear models}
\begin{figure}[!h]
\begin{center}
        \includegraphics[width=1\textwidth]{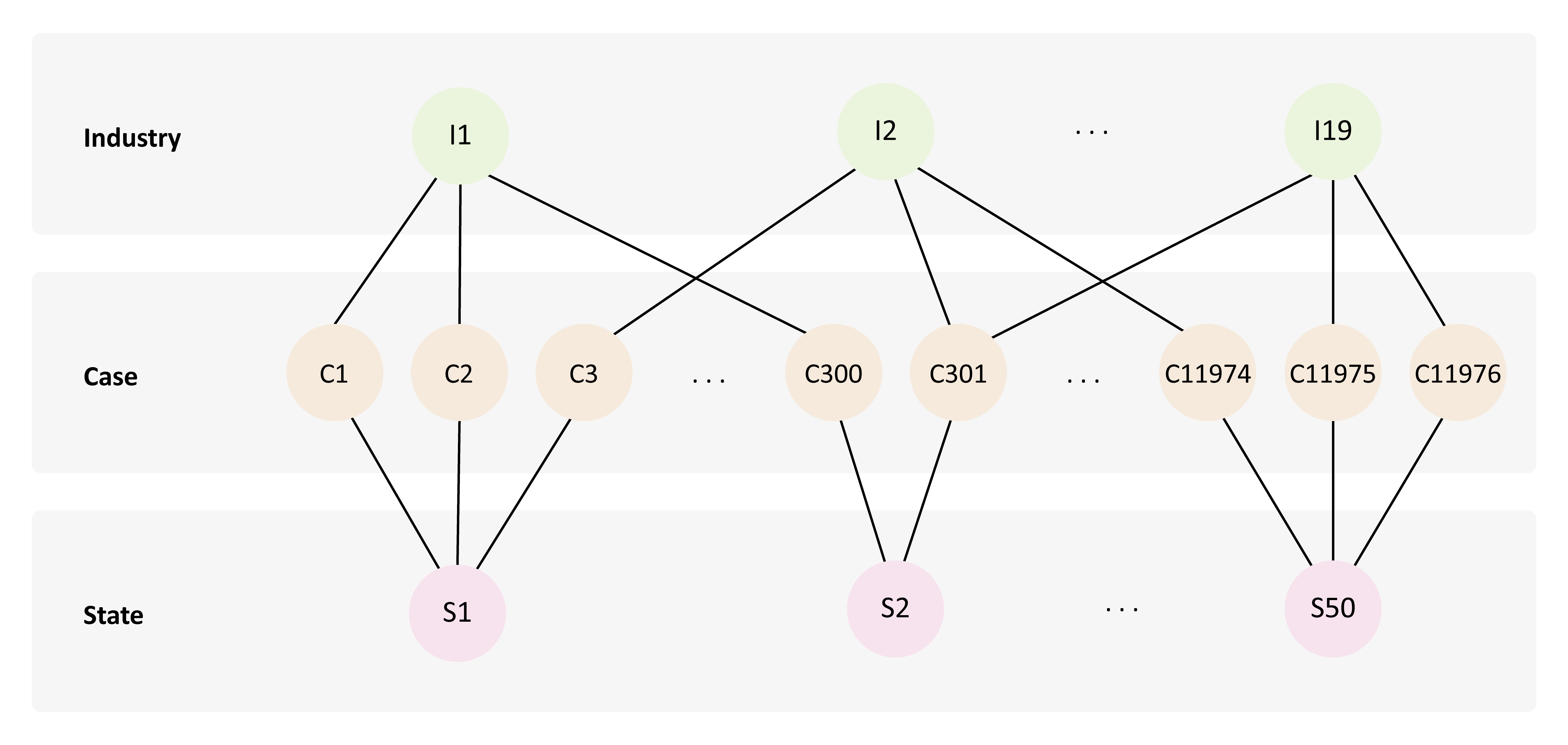}
\caption{{\bf Cross-classified Data.} The multilevel diagram for the cross-classification of two levels of study.}
\label{fig3}
\end{center}
\end{figure}

\quad We explored using ZINB linear models to analyze H-2A violation counts to study the effects of different factors. These factors were explained in detail in Data section. Our data exploration revealed the presence of multiple groups in the dataset, including state and industry. However, we also found that these groups are not hierarchically nested within each other, but instead, they cross~\cite{de2008handbook}. This can be presented schematically for different cases contained within the cross-classification of 50 states by 19 industries, as in Fig~\ref{fig3}. Cases are at the individual level (level 1), and two higher levels are industry and state (level 2). 

\quad Based on the structure of our data, we employed a multilevel zero-inflated negative binomial linear model (MZINBLM) with two crossed random effects -- state and industry in both the conditional and ZI parts of the model. We compared the performance of this model with other alternatives, such as the zero-inflated negative binomial linear model (ZINBLM) without accounting for state effect or industry effect, MZINBLM with only one level (accounting for either state effect or industry effect but not both), and MZINBLM with two levels (accounting both state and industry) in the conditional model. We evaluated the models by comparing AIC, BIC, and likelihood ratio tests. The findings reveal that the MZINBLM with two crossed random effects in both parts of the model (e.g., conditional and ZI) provided a superior fit to the data as compared to the other models considered, see Table~\ref{table4}. 
\begin{table}[!ht]
\caption{{\bf Model Selection}}
\begin{tabular}{c|lllll}
\hline
\bf Model &\bf  \shortstack{Random Effects\\ (Cond Model)} &\bf \shortstack{Random Effects \\(ZI Model)} &\bf AIC & \bf BIC & \bf logLik\\ 
\thickhline
NB     & -     & - & 30543.4 & 30624.7 & -15260.7 \\ 
ZINB  & -     & - & 28804.4 & 28959.6 & -14381.2 \\ 
MZINB1  & state & - & 28545.2 & 28707.8 & -14250.6\\ 
MZINB2  & state \& industry & - & 28423.6 & 28593.6 & -14188.8\\ 
MZINB3  & state \& industry & state & 28031.6 & 28209 & -13991.8\\ 
MZINB4  & state \& industry & state \& industry & 27812.5 & 27997.3 & -13881.3\\ \hline
\end{tabular}
\label{table4}
\end{table}

\quad Let $Y_{i(jk)}$ represent the count outcome for the $i$-th case, belonging to the $j$-th industry group and $k$-th state group, $\mu_{i(jk)}$ be the conditional mean of $Y_{i(jk)}$, and $X_{i1}, \ldots, X_{ip}$ be the $p$ predictor variables at the case level. The MZINBLM formulation is as follows~\cite{fieldingCrossclassifiedMultipleMembership2006}:\\

Level One:
\begin{align*}
\log \big[\frac{\Phi_{i(jk)}}{1- \Phi_{i(jk)}}\big] &= a_{i(jk)} + a_1 X_{i1} + \cdots + a_p X_{ip}  \\
\log(\mu_{i(jk)}) &= \beta_{i(jk)} + \beta_1 X_{i1} + \cdots + \beta_p X_{ip}+ \epsilon_{i}\\
\epsilon_{i} &\sim \mathcal{N}(0, \sigma^2_\epsilon) \\
\end{align*}

Level Two:
\begin{align*}
a_{i(jk)} &= a_{0} + w^{(1)}_j + w^{(2)}_k \quad \text{for} \quad i= 1,2,..., 11976 \\
\beta_{i(jk)} &= \beta_{0} + u^{(1)}_j + u^{(2)}_k\ \quad \text{for} \quad i= 1,2,..., 11976 \\
u^{(1)}_j  &\sim \mathcal{N}(0, \sigma^2_{u^{(1)}}) \\
u^{(2)}_k &\sim \mathcal{N}(0, \sigma^2_{u^{(2)}}) \\
w^{(1)}_j &\sim \mathcal{N}(0, \sigma^2_{w^{(1)}}) \\
w^{(2)}_k &\sim \mathcal{N}(0, \sigma^2_{w^{(2)}}) \\
\end{align*}
Where $a_0, \ldots, a_p$ and $\beta_0, \ldots, \beta_p$ are the corresponding regression coefficients for the logit model and conditional model, respectively; $w^{(1)}_j$, $u^{(1)}_j$ and $w^{(2)}_k$, $u^{(2)}_k$ are the random effects of intercept for state and industry respectively, and $\epsilon_{i}$ is the residual error. The variance of the random effects are denoted by $\sigma^2_{u^{(1)}}$, $\sigma^2_{u^{(2)}}$, $\sigma^2_{w^{(1)}}$, and $\sigma^2_{w^{(2)}}$. 

\quad The results of this study were obtained using R 4.2.2 with glmmTMB package for multilevel zero-inflated generalized linear models. It includes the conditional model, which is the negative binomial distribution, and the zero-inflated model~\cite{brooks2017glmmtmb}. 

\section*{Results} \label{result}
\quad As per the methodology, several models were fit, and their performance was evaluated based on AIC, BIC, and likelihood tests. The best model was selected by comparing the models using these criteria, considering the balance between the goodness of fit and model complexity. 
We used a multilevel zero-inflated generalized linear model to analyze the count of H-2A violations, taking into account fixed effects such as the average size of farms, number of small/medium establishments, labor intensity, average net income per farm, number of task forces, length of alleged violations, poverty rate, fatal injury rate, and minimum standard wages, as well as crossed random effects from state and industry. 

\quad The selected model identified the significant factors affecting the count outcome. Table~\ref{table5} presents the results of the selected MZINBLM, where the first part pertains to the conditional model. This part describes the correlation between the independent variables and the logarithm of the count data, which can be interpreted as the change in the expected H-2A violation counts associated with a unit increase in the corresponding variable while holding all other variables constant. The second part follows that corresponds to the zero-inflation model. This includes log odds of count outcome (estimate) for predicting excess zeros and their corresponding p-values.

\begin{table}[!ht]
\caption{{\bf The Multilevel Zero-inflated Negative Binomial Linear Model}}
\begin{tabular}{lcl}
\hline
\bf Factor &\bf Estimate  &\bf Pr($>|z|$) \\ 
\thickhline
\textit{Conditional Model}               &  &\\
Intercept                                & 2.407  & $<$ 2e-16***\\
Average size of farms                    & 0.122  & 0.006**\\ 
Number of small/med establishments       & 1.369  & 0.000***\\ 
Labor intensity                          & -0.030 & 0.921\\ 
Average net income per farm              & -0.129 & 0.185\\ 
Number of task forces                    & -0.017 & 0.559\\ 
Length of alleged violations             & 0.123  & 0.307\\ 
Poverty rate                             & 0.253  & 0.008**\\ 
Fatal injury rate                        & -0.109 & 0.169\\ 
Minimum standard wages                   & -0.082 & 0.159\\ 
\textit{Zero-inflated Model}            &  &\\
Intercept                                & 2.829  & 0.000***\\
Average size of farms                    & -0.125 & 0.247\\ 
Number of small/med establishments       & -0.215 & 0.821\\ 
Labor intensity                          & -2.386 & 0.039*\\ 
Average net income per farm              & 0.025  & 0.838\\ 
Number of task forces                    & -0.066 & 0.026*\\ 
Length of alleged violations             & -1.585 & $<$ 2e-16***\\ 
Poverty rate                             & 1.065  & $<$ 2e-16***\\ 
Fatal injury rate                        & -0.725 & 2.74e-07***\\ 
Minimum standard wages                   & -0.441 & 1.43e-09***\\ \hline
\end{tabular}
\begin{flushleft}  Significant codes: 0 ***, 0.001 **, 0.01 *
\end{flushleft}
\label{table5}
\end{table}

\quad  The results indicate that the intercept, average size of farms,  the number of small/medium establishments, and the poverty rate have a significant correlation with the mean H-2A violation counts in the conditional model. Additionally, the zero-inflated model suggests several factors are significantly associated with the outcome, including labor intensity, number of task forces, length of alleged violations, poverty rate, fatal injury rate, and minimum standard wages. Notably, the average net income per farm is not a significant factor, indicating no significant relationship between this variable and the count outcome. 

\quad Upon analyzing the regression results, we noted a positive correlation between the significant factors and the target variable in the conditional model. However, it is essential to note that correlation does not necessarily indicate causation. This relationship may not imply causality due to potential confounding factors, reverse causality, or random variation in the data. Therefore, interpreting these results requires caution, and further research with rigorous study designs and control for confounding factors is needed to establish a causal relationship between these variables. Specifically, a one-unit increase in average farm size correlates with $e^{0.122} = 1.13$ increase in the mean H-2A violation counts. Furthermore, the positive coefficient for the number of small/medium establishments suggests that DOL identified more H-2A violations in states or industries with higher numbers of small/medium establishments (less than 20 employees). The poverty rate is significant in conditional and ZI models. It illustrates that states with higher poverty rates have higher mean H-2A violation counts. According to the ZI model, an increase in the poverty rate is associated with a higher likelihood of structural zero, where no violations are discovered in organizations that are not at risk of H-2A violations. Therefore, it is more likely that zero counts for states with high poverty rates happen due to one of two reasons. First, DOL investigated other violations (not H-2A). Second, the organization's structure does not require hiring H-2A workers, making it impossible for such organizations to have H-2A violations.

\quad The ZI model also highlights other significant factors such as labor intensity, the number of task forces, the length of alleged violations, the fatal injury rate, and minimum standard wages. The ZI model presents a negative correlation between these predictors and the log of odds ($log(\frac{\Phi}{1-\Phi})$), indicating that increasing one of these predictors is associated with less likelihood of excessive zero (structural zero). The states with higher labor intensity, task forces, length of alleged violations, fatal injury rates, and minimum standard wages are more likely to have nonstructural zeros. Zero H-2A violation counts in these states could happen for two reasons. Although they are at risk of violations, DOL could not detect them. The following reason is that no H-2A violations occurred during the investigation period. Therefore, enhancing inspection methods in these states could lead to more effective detection of labor trafficking.

\section*{Discussion} \label{discussion}
\quad This paper presented a multilevel zero-inflated negative binomial regression model to handle cross-classified data with excessive zeros, particularly regarding labor violations in agriculture.  We employed a multilevel regression model to investigate the association between different factors and the frequency of H-2A labor violations. The study emphasizes the need for a flexible labor inspection strategy that considers the location and type of agricultural industry.

\quad The ZI model revealed several essential factors correlating with the probability of excessive zeros. Our analysis of DOL's investigations showed that the reasons for zero H-2A violation counts vary depending on the location and the type of agricultural industry. The ZI model's insights suggest that zero counts in industries with higher labor intensity are more likely to occur either because the DOL could not detect the violations or no violation occurred during the investigation. Similar relationships can be observed for other factors, such as the number of task forces, length of alleged violations, fatal injury rate, and minimum standard wages. This study offers valuable insights for labor inspectors and agencies involved in H-2A labor inspections. These findings can guide them in making better-informed decisions regarding resource allocation, encompassing aspects such as budget, the number of inspectors, and inspection time. By doing so, they can enhance the effectiveness and efficiency of resource distribution, particularly in areas with a higher risk of H-2A violations. 

\quad This study has several limitations that should be considered. First, while the regression analysis identified correlations between certain factors and H-2A violation counts, it did not establish causal relationships. Future research should aim to determine the causal effects of these factors on labor violations in order to better understand the underlying dynamics. Second, the absence of a comprehensive tracking system for farm workers' data, including health issues, access to healthcare, income, housing conditions, and demographic information, limited our dataset construction. As a result, we relied on available data sources. This constraint suggests that our findings should be interpreted carefully, and additional, more extensive research is necessary to gain a deeper understanding of H-2A violations in the agricultural sector. It is important to acknowledge that while the DOL plays a role in identifying violations, there is still room for improvement in effectively uncovering abuses. There are numerous reasons why H-2A workers might not report abuses or concerns, and previous research has consistently highlighted the need for more proactive efforts to uncover such abuses. Furthermore, the WHD data we obtained did not clarify whether the DOL conducted H-2A violation inspections in cases with zero counts. This leaves the possibility of undetected or unreported cases, which may have led to an underestimation of H-2A violations in specific states and industries. We attempted to address this issue by using a zero-inflated model to account for excessive zeros. However, this highlights the need for improved data collection strategies to better track labor inspections and accurately assess the prevalence of H-2A labor violations. 

\quad In addition to the insights that can be gained in this study, the use of H-2A violation data may also provide critical insights into labor trafficking, which is significantly under identified.  Though the data is not perfect, it shows the potential for administrative data related to workplace safety and to help inform our understanding of places and situations where workers may be vulnerable to abuses and are in need of protection.

\section*{Conclusion} \label{conclusion}

\quad In conclusion, our findings have significant implications for law enforcement and labor inspection agencies in crafting labor inspection strategies based on the characteristics of each state and industry. By pinpointing significant factors correlated with labor violations, such as larger acreage farms, establishments with less than 20 employees, and higher poverty rates, labor inspectors can refine their approaches to target these areas more effectively. Furthermore, it is essential to recognize that the absence of H-2A violations in certain states or industries does not necessarily imply that no violations have occurred. Our multilevel zero-inflated modeling approach can assist in identifying and addressing labor violations more efficiently, suggesting that current labor inspection strategies may require improvements to detect H-2A labor violations more effectively.

\section*{Acknowledgment} \label{acknowledgment}

\quad We thank Aubrey Sneesby, Chase Childress, Cao Jainfei, Elaine Klatt, Joseph Pak, and Margaret Clark for their thoughtful comments and assistance related to gathering data, data cleaning, and methodology.

\bibliography{References}
\end{document}